\documentclass[twocolumn,showpacs,preprintnumbers,amsmath,amssymb]{revtex4}

\usepackage{graphicx}
\usepackage{dcolumn}
\usepackage{bm}

\begin{document}


\title{Computation of Resistive Wakefields}

\author{Adina Toader and Roger Barlow}
 \altaffiliation[Also at ]{The Cockcroft Institute, Daresbury Science and Innovation Campus, Warrington, WA4 4AD, United Kingdom}
 \email{Adina.Toader@manchester.ac.uk, Roger.Barlow@manchester.ac.uk}
  \affiliation{%
  School of Physics and Astronomy, The University of Manchester, Manchester M13 9PL, United Kingdom}%
\begin{abstract}
We evaluate longitudinal resistive wakefields for cylindrical beam
pipes numerically and compare the results with existing
approximate formul\ae. We consider an ultra-relativistic bunch
traversing a cylindrical, metallic tube for a model in which the
wall conductivity is taken to be first independent and second
dependent on frequency, and we show how these can be included
simply and efficiently in particle tracking simulations. We also
extend this to higher order modes, and to the transverse wakes.
This full treatment can be necessary in the design of modern
nano-beam accelerators.
\end{abstract}

\maketitle

\section{\label{sec:level1}Introduction}
The requirements of modern accelerators are placing increasing
demands on technology, with small precisely-defined beam bunches
passing through physically small beam pipes and collimators. This
means that the effects of wakefields produced by induced charges
and currents are becoming increasingly important and require
accurate calculation to ensure they do not dilute the emittance of
the beam.

The wakefields from particles in a bunch may affect other
particles in the same bunch (intrabunch wakes) and also particles
in subsequent bunches (interbunch wakes). For physically small
apertures the effects of short-range intra bunch forces become
important and need to be considered, as well as the more familiar
long range interbunch effects.

For a relativistic particle in a perfectly conducting uniform beam
pipe the wake is zero. For a non-uniform beam pipe of finite
conductivity the effect is thus separated into geometric and
resistive wakes, the second of which is considered here.

The wake is due to the Lorentz force
$\vec{F}=q(\vec{E}+~\vec{v}\wedge\vec{B})$, however for the
longitudinal component $F_{\parallel}$ the second term is zero and
only $E_{\parallel}$ need be considered. For a purely resistive
wake, the wake field, $\vec{E}+~\vec{v}\wedge\vec{B}$, is constant
and the integrated effect of the complete transit of the bunch
through a section of pipe, generally called the wake potential, is
obtained just by multiplying the wake field by the pipe length.
The wake field is found~\cite{Chao} by writing down Maxwell's
Equations in the beam aperture and in the beam pipe and matching
them subject to appropriate boundary conditions. These solutions
can be written as a sum over angular modes. For many purposes only
the knowledge of the leading modes ($m=0$ and $m=1$) is adequate,
however as requirements become more stringent a technique is
needed that will work in the general case.

Finding solutions of Maxwell's equations is more readily done in
frequency (or wavenumber) space rather than physical space, as
differentiation becomes multiplication. The Fourier Transform of
the wake is the impedance

\begin{equation}
\tilde {\vec F}(k)=\int \vec F(s) e^{i k s} ds \qquad \vec F(s)={1
\over 2 \pi} \int \tilde{\vec F}(k) e^{-i k s} dk
\end{equation}
where $s$ is the distance along the beam axis between a leading
particle which creates the field and a trailing (witness) particle
which feels the effect.

For many purposes a knowledge of the impedance suffices. The `kick
factor' averaged over all the bunch is also often useful and
expressions are given in the literature. However the kick factor
only gives the mean effect for the whole bunch and in particle
tracking codes an expression is needed for the physical impulse
that one leading particle has on another trailing particle.
Existing approaches use expressions in different
regimes~\cite{Placet}, where the division is somewhat arbitrary.
We have found an approach which unifies, simplifies and speeds up
the calculations, and makes the underlying physics clear. It
includes short range and long range wakes with no artificial
division between them. It can also be extended to arbitrary
angular order.

Chao~\cite{Chao} gives a general formula for the impedance, from
which he obtains an expression for the physical wake which is
valid in the long-range limit. Bane and Sands~\cite{Bane,
BaneSands} extend this to shorter range though still making
approximations.

Gluckstern, van Zeits and Zotter\cite{GZZ} consider resistive
wakefields for circular, elliptical and rectangular pipes, but
they only evaluate the impedance, not the physical wake, and only
consider the lowest ($m$=0 longitudinal, $m=1$ transverse) modes.
Yokoya~\cite{Yokoya} also considers beam pipes of general cross
section, though his results are self-admittedly complicated and
not easy to implement. Lutman, Vescovi and Craievich~\cite{LVC}
generalise the results to elliptical beam pipes, including
circular and planar apertures as special cases. Although a very
general approach, they consider only leading modes and the
examples they give are specific and not directly applicable to
simulation programs.

\section{\label{sec:level2}The longitudinal wake for $m=0$}
The Fourier Transform of the $m=0$ mode of the longitudinal
component of the wakefield is given by \cite{Bane}
\begin{equation}\label{eq:first}
\tilde E_z(k)={2 q \over b} {1 \over {i k b \over 2} - \left(
{\lambda \over k } + {k \over \lambda} \right) \left( 1 + {i \over
2 \lambda b} \right) }
\end{equation}
where
\begin{equation}
\lambda(k) = \sqrt{2 \pi \sigma |k| \over c} (i+sgn(k))
\label{eqsign}
\end{equation}
with $q$ the charge of the particle, $b$ the radius of the tube,
$\sigma$ the conductivity of the pipe, and $c$ the speed of light.
This assumes axial symmetry, the validity of Ohm's law,
relativistic particles, and that the skindepth is smaller than
both the thickness of the pipe and the tube radius, but is
otherwise general. This enables the Bessel function solution of
$\tilde{\vec{F}}(k)$ to be replaced by a sinusoidal form, i.e. the
asymptotic form of $J_0(x)\propto x^{-1/2}e^{ix}$, as suggested by
Chao(\cite{Chao},p.43), which leads to the
$\left(\frac{i}{2\lambda b} \right)$ term in the
Equation~\ref{eq:first}.

It is convenient to introduce $s_0$, the scaling length
\begin{equation}\label{eq:third}
s_0 = \root 3 \of {c b^2 \over 2 \pi \sigma }
\end{equation}

and thus the dimensionless quantities $K$, the {\it scaled wave
number} and $s'$, the {\it scaled length}
\begin{eqnarray}\label{eq:forth}
K=s_0 k \\
s'={s\over s_0}
\end{eqnarray}

It is useful to note that
\begin{eqnarray}
&& \lambda=\frac{b \sqrt{|K|}}{s_0^2}(1 \pm i) \qquad
\frac{k}{\lambda} = \frac{1}{2} \frac{s_0 \sqrt{|K|}}{b} (1 \mp
i) \nonumber\\
&& \frac{\lambda}{k} = \frac{b}{s_0 \sqrt{|K|}} (1 \pm i)
\label{eq:klambda}
\end{eqnarray}
where the upper sign applies for
positive $k$, the lower for negative $k$.

To find the corresponding wakefield requires the inverse Fourier
transform. This has been much studied in the
literature\cite{Bane,Chao}, using various approximations forms of
Equation~\ref{eq:first} and evaluating them by a contour integral.
By contrast we will do the integration numerically, enabling us to
provide a general technique without making approximations as has
been done previously~\cite{Chao, BaneSands, Placet} and to
evaluate the different regions wherein such approximations are
valid.

The back transform can be written
\begin{eqnarray}\label{eq:Ez}
{E}_z (s)= \frac{1}{2\pi}\int_{-\infty}^{\infty}
\Bigl(Re[f_{even}(k)]\cos(ks)\nonumber\\
+Im[f_{odd}(k)]\sin(ks)\Bigr)dk  \\
{E}_z (s)= \frac{s_0}{2\pi}\int_{-\infty}^{\infty}
\Bigl(Re[f_{even}(K)]\cos(Ks')\nonumber\\
+Im[f_{odd}(K)]\sin(Ks')\Bigr)dK
\end{eqnarray}
where the functions are the even and odd parts of
$\tilde{E}_z(k)$. This separation is necessary to avoid problems
with the alternating signs and the modulus operations in Equations
\ref{eq:klambda}.

\subsection {The first order approximation}
In the limit of large $b$ compared to $s_0$ and neglecting high
and low frequencies Equation~\ref{eq:first} can be approximated by
\begin{equation}\label{eq:sixth}
\tilde E_z (k)= -{2 q k\over \lambda  b}
\end{equation}
The Fourier Transform is given by Chao~\cite{Chao} as
 \begin{equation}\label{Chao}
 E_z(s)={q
\over 2 \pi b} \sqrt{c \over \sigma} s^{-{3 \over 2}}.
\end{equation}
However we can also evaluate it numerically, in preparation for
more complicated forms of  $\tilde E_z(k)$. We have
\begin{eqnarray}\label{eq:seventh}
f_{even}(K)&=&\frac{1}{2}\Bigl[f(K)+f(-K)\Bigl]=-\sqrt K, \nonumber \\
f_{odd}(K)&=&\frac{1}{2}\Bigl[f(K)-f(-K)\Bigl]=i \sqrt K
\end{eqnarray}
where the functions are the even and odd parts of $\tilde E_z(k)$,
having taken out a common factor of ${q \over b^2}$. The even part
is purely real and the odd part is purely imaginary, as they must
be.

A function like $\sqrt K\sin(K)$ presents problems for numerical
integrals as the function is oscillating with increasing
amplitude, and any summation technique (trapezoidal rule,
Simpson's rule or Gaussian quadrature) will not work. However we
know that the integral exists because it can be done analytically.
We can perform the $K$ integral by first integrating with respect
to $s'$, thereby dividing by a factor of $K$ so that the
oscillations decrease in size and the numerical integration can
succeed
\begin{eqnarray}\label{eq:eigth}
\int_0^s E_z(x)dx &=& s_0 \int_0^{s/s_0} E_z(x') dx' \nonumber \\
&=& {s_0 q \over \pi b^2} \int_0^\infty \Bigl(
\frac{Re[f_{even}(K)]}{K} \sin{(Ks')} \nonumber \\
&-& \frac{Im[f_{odd}(K)]}{K} \cos{(Ks')}\Bigr)dK
\end{eqnarray}

The result can then be differentiated numerically, using an
intermediate interpolating function, to give the desired
wakefield.

\begin{figure}[!htb]
  \begin{center}
   \includegraphics[scale=0.73,angle=0]{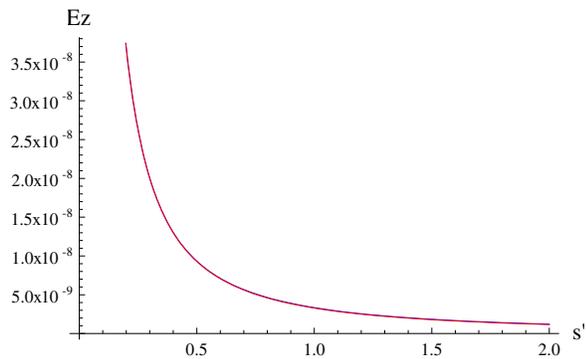}
   \caption{The long range wake given by Equation~\ref{eq:eigth} and Equation~\ref{Chao}~\cite{Chao}.}
   \label{fig:Fig1}
  \end{center}
 \end{figure}

This can be verified and results are shown in Fig.~\ref{fig:Fig1},
for $b=1$ cm and $\sigma=5.8\times 10^{7}\,(\Omega m)^{-1}$. These
results are indistinguishable from those obtained by plotting
Chao's formula, Equation~\ref{Chao}.

\subsection {A more accurate formula}
If the large $s$ requirement is relaxed then a second term has to
be included in the denominator and Equation~\ref{eq:first} is
approximated by
\begin{equation}\label{BandS}
\tilde E_z(k)=\frac{2 q}{b} \frac{1}{\frac{i k b}{2} -
\frac{\lambda}{k}}
\end{equation}
The even and odd parts are
\begin{eqnarray}
  f_{even}(K) &=& - \frac{\frac{2}{\sqrt{K}}}{\Bigl(\frac{K}{2}-\frac{1}{\sqrt{K}}\Bigr)^2
  +\frac{1}{K}},\nonumber \\
  f_{odd}(K) &=&
-\frac{2i
\Bigl(\frac{K}{2}-\frac{1}{\sqrt{K}}\Bigr)}{\Bigl(\frac{K}{2}-\frac{1}{\sqrt{K}}\Bigr)^2+\frac{1}{K}}
\end{eqnarray}

\begin{figure}[!htb]
  \begin{center}
   \includegraphics[scale=0.75,angle=0]{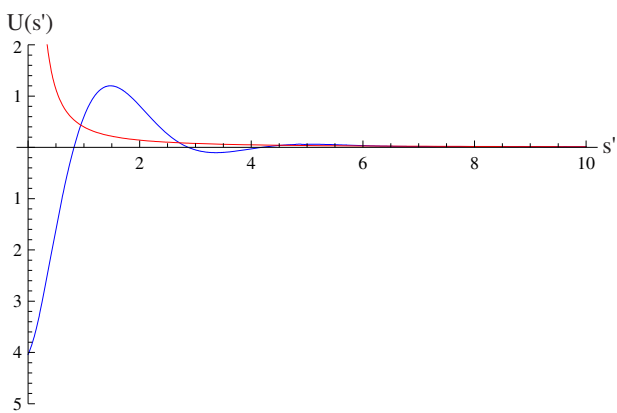}
   \caption{ The longitudinal wake in the second
approximation (blue line) with the long range form (red line) for
comparison.}
   \label{fig:Fig2}
  \end{center}
 \end{figure}
Although the wake is a function of three parameters ($s, b$ and
$\sigma$), the use of the scaling length $s_0$ enables it to be
written as a universal function $U(s')$, where $E_z(s,b)=~{s_0 q
\over b^2} U(s/s_0)$. The result is  shown in Fig.~\ref{fig:Fig2},
together with the lowest order approximation. They differ greatly,
although above $s'\simeq 6$ the agreement is actually quite good
and Equation~\ref{Chao} ia a good approximation for long range
wakes. This transform can also be performed analytically, using
contour integration as was done by Bane and Sands~\cite{Bane} and
our Fig.~\ref{fig:Fig2} agrees with their results, including the
key value $U(0)=-4$.

\subsection{The full formula}
For the full version of Equation~\ref{eq:first} one gets
\begin{eqnarray}\label{full}
&&f_{even}(K)= \nonumber \\
&&\frac{-8 \Bigr(\xi^2+2\xi\sqrt{K}+\frac{4}{\sqrt{K}}\Bigl)}
{4\Bigr[\xi\sqrt{K}-\frac{1}{K}\Bigr( \xi+2\sqrt{K} \Bigl)+K
\Bigl]^2+
\Bigr(\xi^2+\xi 2\sqrt{K}+\frac{4}{\sqrt{K}}\Bigl)^2} \nonumber\\
\newline
&&f_{odd}(K)= \nonumber \\
&&\frac{-16i \Bigr[\xi\sqrt{K}-\frac{1}{K}\Bigr( \xi+2\sqrt{K}
\Bigl)+K \Bigl]} {4\Bigr[\xi\sqrt{K}-\frac{1}{K}\Bigr(
\xi+2\sqrt{K} \Bigl)+K \Bigl]^2+ \Bigr(\xi^2+\xi
2\sqrt{K}+\frac{4}{\sqrt{K}}\Bigl)^2} \nonumber\\
\end{eqnarray}
where we have introduced the dimensionless quantity $\xi
=s_0^2/b^2$. Although this is no longer a universal curve, it can
still be expressed as a function of two variables ($s'$ and $\xi$)
rather than the full set of three. The earlier approximation
corresponds to the function at $\xi=0$.
\begin{figure}[!htb]
  \begin{center}
   \includegraphics[scale=0.75,angle=0]{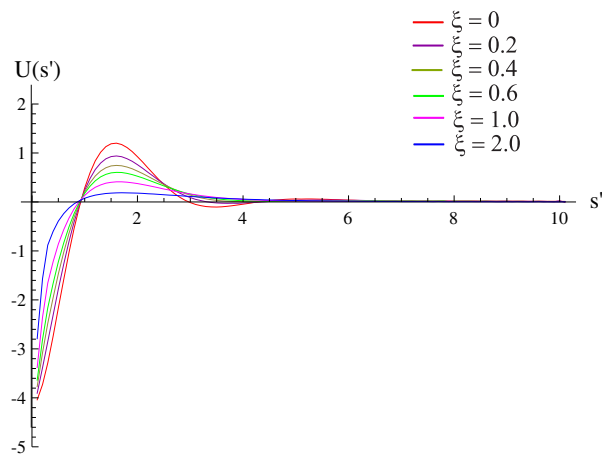}
   \caption{ The exact longitudinal range wake for
various values of $\xi$.}
   \label{fig:Fig3}
  \end{center}
 \end{figure}

Fig.~\ref{fig:Fig3} shows how the function when evaluated with our
numerical technique changes for different values of $\xi$. It can
be seen that for values of below about $0.1$ the approximation is
very good. For a copper beam pipe with a radius of $1$ cm the
scaling length is of order $20$ microns, so $\xi$ is very small in
all practical cases at present. However, for possible future
collimators with very low conductivity and small radius it might
need to be considered. It is also important to check that the
simpler formula is valid before using it.

\section {Longitudinal: Higher order  modes}\label{sec:HOM}

\begin{figure}[!htb]
  \begin{center}
   \includegraphics[scale=0.75,angle=0]{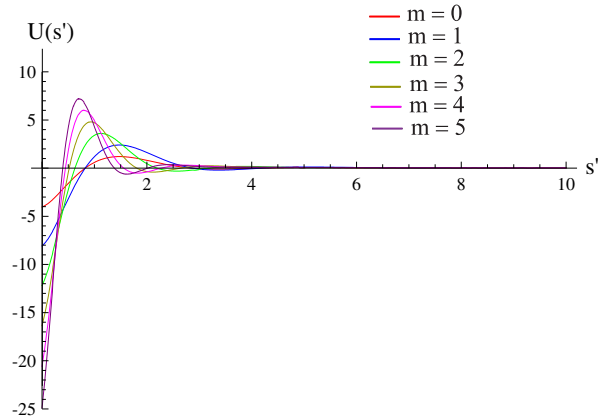}
   \caption{The longitudinal wake for various modes.}
   \label{fig:Fig4}
  \end{center}
 \end{figure}

For higher modes, using the same technique of matching the
solutions of Maxwell's equations in the aperture and the pipe, and
using the same approximations as for Equation~\ref{BandS}  for
$m>0$
\begin{equation}\label{Ezm_BS}
  \tilde{E}_z^m(k) = \frac{2 I_m}{b^{2m+1}}\frac{1}{\frac{i k
  b}{m+1}-\frac{\lambda}{k}}
\end{equation}
where $I_m$ is the charge moment of order $m$: \(\int\int
\rho(r,\theta) r^m \cos(m \theta)\,dr \,d\theta \). Any angular
distribution of charges at a particular radius can be described in
terms of these moments. Equation~\ref{Ezm_BS} can be separated
into odd and even parts, taking out a factor of \(I_m
s_0/b^{2m+2}\) as before for simplicity
\begin{eqnarray}
f_{even}(K) &=& -
\frac{\frac{2}{\sqrt{K}}}{\Bigl(\frac{K}{m+1}-\frac{1}{\sqrt{K}}\Bigr)^2
  +\frac{1}{K}} \nonumber\\
f_{odd}(K) &=& -
\frac{2i\Bigl(\frac{K}{m+1}-\frac{1}{\sqrt{K}}\Bigr)}{\Bigl(\frac{K}{m+1}-\frac{1}{\sqrt{K}}\Bigr)^2+\frac{1}{K}}
\end{eqnarray}
Results are shown in Fig.~\ref{fig:Fig4}. It can be seen that
there are quite large differences in behaviour, even though only a
few terms in the formula contain an $m$. The value at $s=0$ is
given by $U^m(0)=-4(m+1)$.

The full formula for higher modes, corresponding to Equation
\ref{eq:first}, is
\begin{equation}\label{fullm}
  \tilde{E}_z^m(k) = \frac{4 I_m}{b^{2m+1}}\frac{1}{\frac{i k
  b}{m+1}-\Bigr(\frac{2k}{\lambda}+\frac{\lambda}{k}\Bigl)
  \Bigr(1+\frac{i}{2\lambda b}\Bigl)-\frac{i m}{kb}}
\end{equation}
(Note that Equation~\ref{fullm} is valid only for $m>0$.)
\begin{widetext}
This can be separated into odd and even parts as for Equation
\ref{full}
\begin{eqnarray}
f_{even}(K) &=& - 8\frac{\xi^2+2\xi \sqrt{K}+\frac{4}{\sqrt{K}}}{
 4\Bigr[\xi \sqrt{K}-\frac{1}{K}(
\xi+2\sqrt{K})+ 2\Bigr(\frac{K}{m+1}-\xi \frac{m}{K}
\Bigl)\Bigl]^2+
\Bigr(\xi^2+\xi 2\sqrt{K}+\frac{4}{\sqrt{K}}\Bigl)^2} \nonumber\\
f_{odd}(K) &=& - 8i \frac{2\Bigr[\xi \sqrt{K}-\frac{1}{K}(
\xi+2\sqrt{K})+ 2\Bigr(\frac{K}{m+1}-\xi \frac{m}{K}
\Bigl)\Bigl]}{4\Bigr[\xi \sqrt{K}-\frac{1}{K}( \xi+2\sqrt{K})+
2\Bigr(\frac{K}{m+1}-\xi \frac{m}{K} \Bigl)\Bigl]^2+
\Bigr(\xi^2+\xi 2\sqrt{K}+\frac{4}{\sqrt{K}}\Bigl)^2}
\end{eqnarray}
\end{widetext}

\begin{figure}[!htb]
  \begin{center}
   \includegraphics[scale=0.55,angle=0]{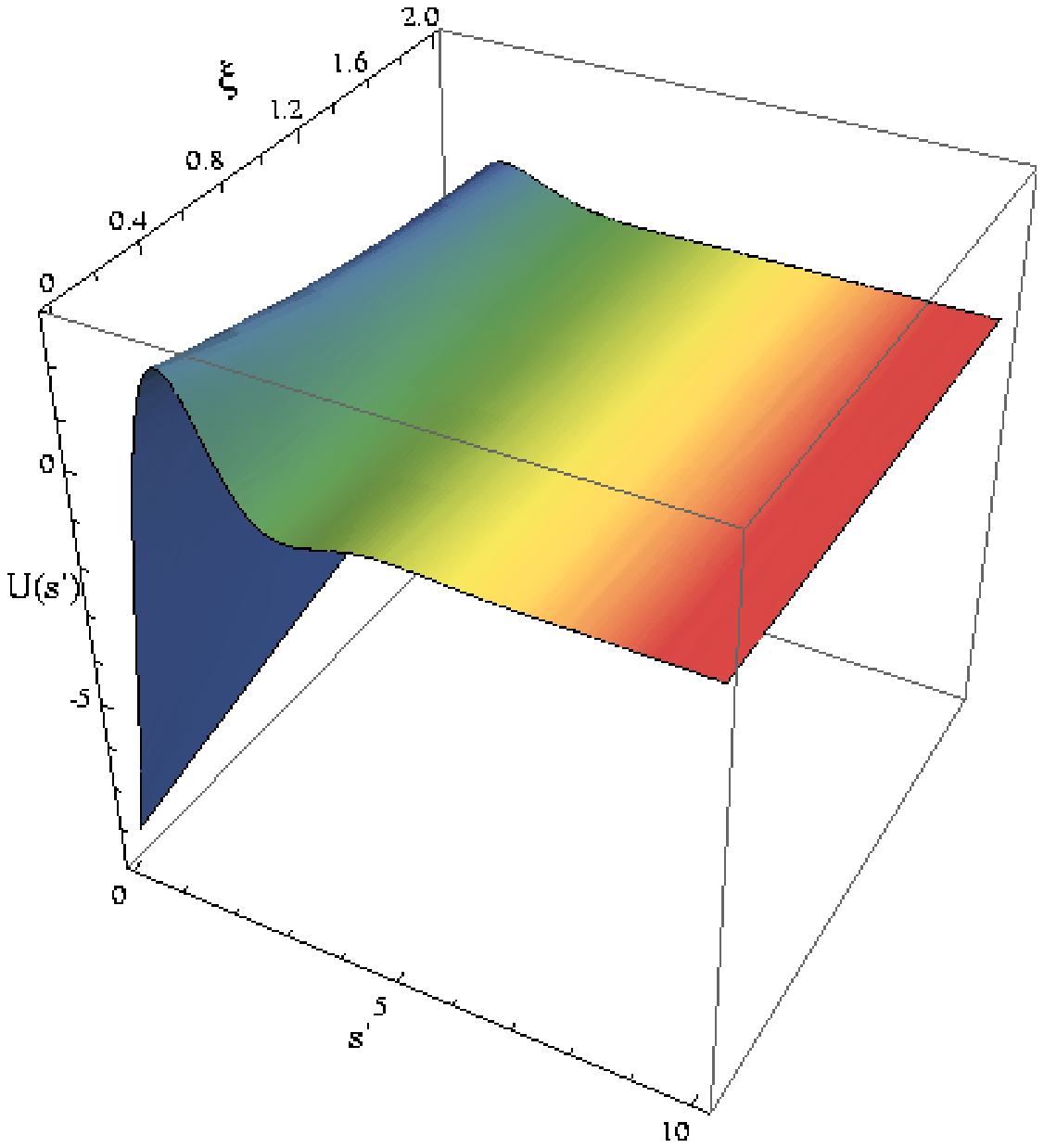}
   \includegraphics[scale=0.55,angle=0]{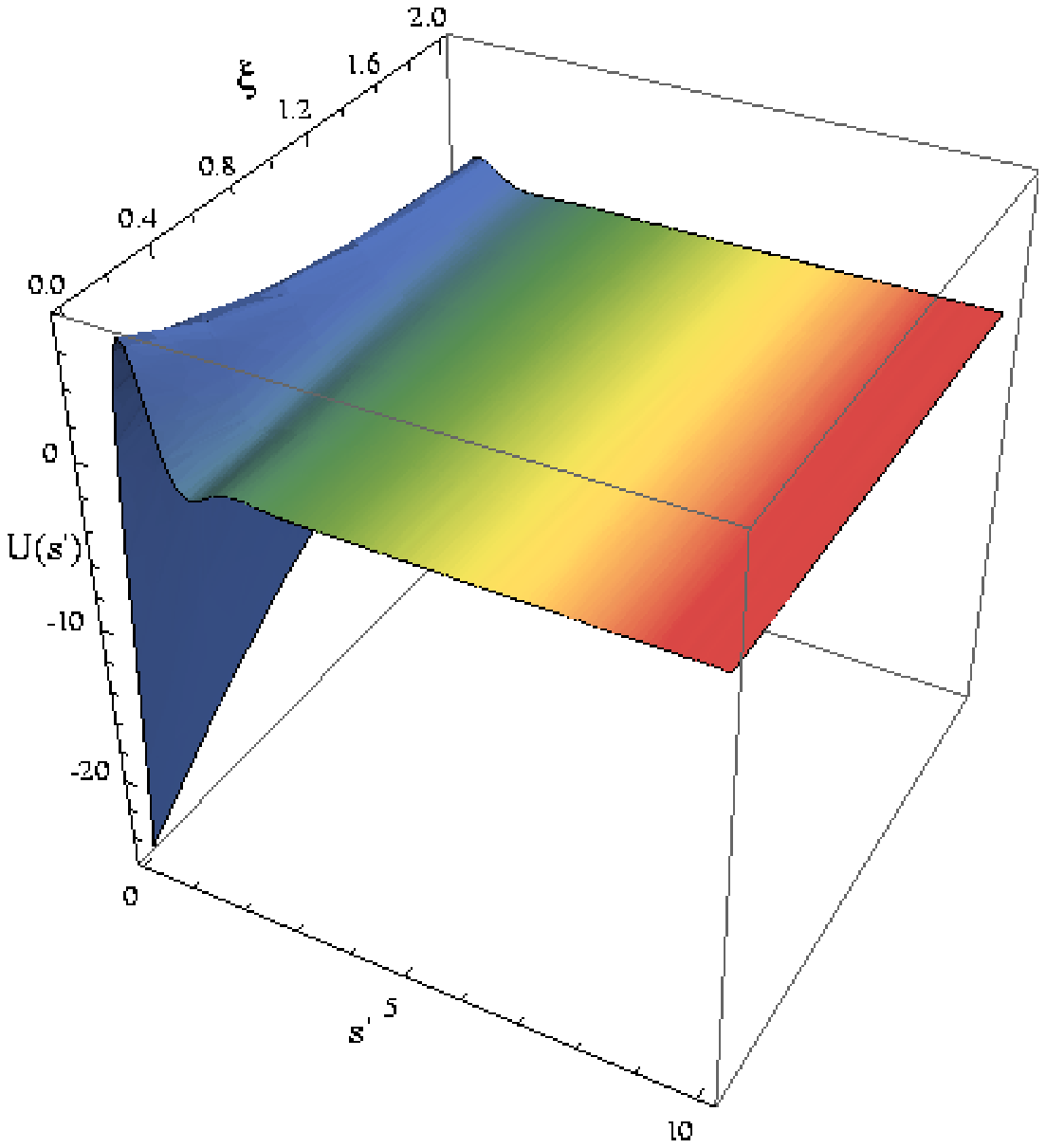}
   \caption{ The $m=1$ (top) and $m=5$
(bottom) wake as a function of $s'$ and $\xi$.}
   \label{fig:Fig5}
  \end{center}
 \end{figure}
We show the dependence on $\xi$ in Fig.~\ref{fig:Fig5} for $m=1$
and $m=5$. It can be seen that the dependence on $\xi$ increases
for higher modes. However even for the $m=5$ mode shown the values
of $\xi$ that would correspond to any reasonable beam pipe
diameter and conductivity are still so small that the deviation is
probably unimportant.

It is frequently stated that the $m=1$ wakefield is related to the
$m=0$ wake by a factor ${2 \over b^2}$. From Fig.~\ref{fig:Fig4}
it can be seen that the first two angular modes do indeed have the
same shape, although higher modes are significantly different. The
equality $E^1={2 \over b^2} E^0$ is true for Equations \ref{BandS}
and \ref{Ezm_BS}, as can be see by making the substitution.

\begin{figure}[!htb]
  \begin{center}
   \includegraphics[scale=0.75,angle=0]{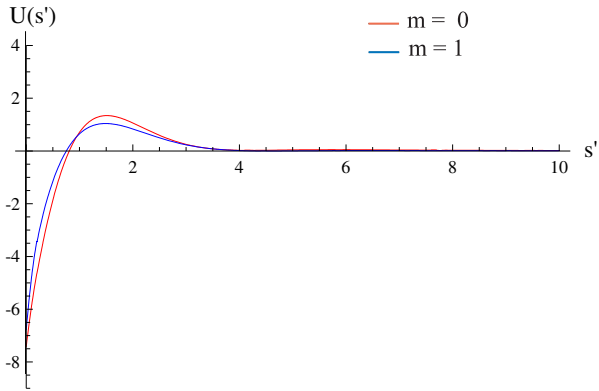}
   \caption{ The $m=0$ longitudinal wake, doubled, and the $m=1$ wake for
$\xi=0.5$.}
   \label{fig:Fig6}
  \end{center}
 \end{figure}

However this is not true for more general formulae, Equations
~\ref{fullm} and ~\ref{full}, as is shown in Fig.~\ref{fig:Fig6}.
For $\xi=~0.5$, which is admittedly large, there is a clear
difference between the two curves.

\section {\label{sec:level3}Transverse wakes}
Transverse wake effects are generally of more importance than
longitudinal wakes, especially for collimator studies. The
transverse wakefield experienced by a particle with transverse
position $r$ due to another particle at $r'$ can be written as a
sum over angular modes
\begin{equation}\label{eq:transverse}
\vec F_T(r,\theta,s)=\sum_m r^{m-1} r'^{m}(\hat r
cos(m\theta)-\hat\theta sin(m\theta)) W^m_T(s)
\end{equation}
where $\theta$ is the angle between the two particles and $s$ is
the distance between them. The property \ref{eq:transverse}
applies to the force components and not to the electromagnetic
field components(\cite{Chao}, p.56), and for the transverse wakes
the magnetic part has to be included.

The Panofsky-Wenzel theorem
\begin{equation}
\nabla_T F = {\partial \vec F_T \over \partial z}
\end{equation}
applies term by term giving
\begin{equation} {W'}^m_T(s)=E_z^{m}(s).
\end{equation}

Thus the transverse wake at any order can be obtained by
integrating the longitudinal wake. This is especially convenient
for our method as this integral is what is calculated in the
inverse transform.
\begin{figure}[!htb]
  \begin{center}
   \includegraphics[scale=0.7,angle=0]{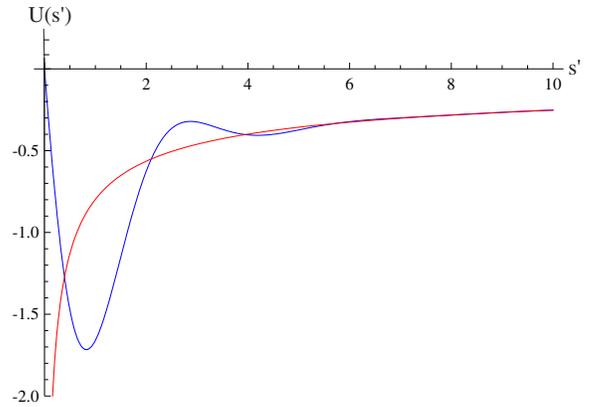}
   \caption{ The Transverse wake in the first
and second approximation (red and blue curves respectively).}
   \label{fig:Fig7}
  \end{center}
 \end{figure}
\begin{figure}[!htb]
  \begin{center}
   \includegraphics[scale=0.75,angle=0]{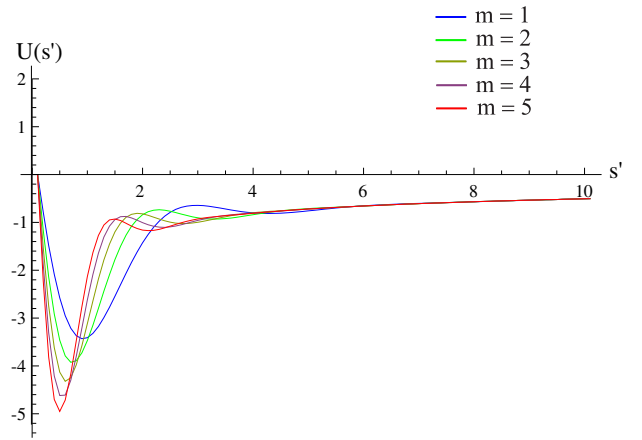}
   \caption{ Transverse wakes - various modes with $\xi$=0.}
   \label{fig:Fig8}
  \end{center}
 \end{figure}
 \begin{figure}[!htb]
  \begin{center}
   \includegraphics[scale=0.75,angle=0]{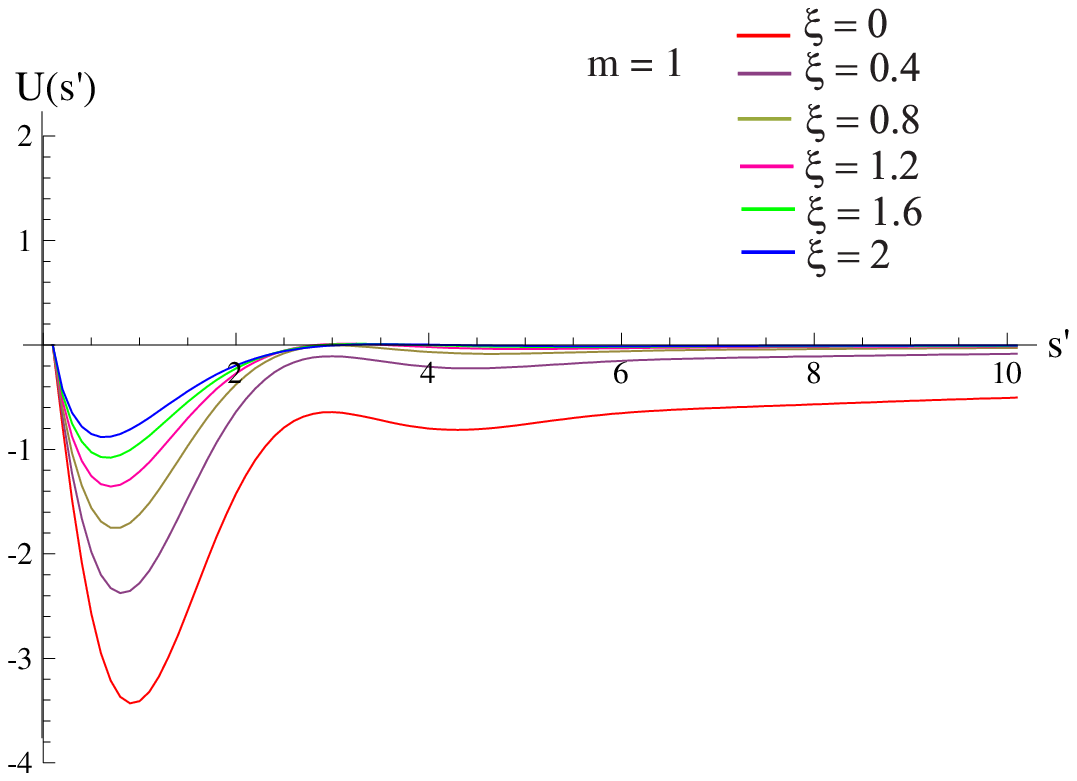}
    \includegraphics[scale=0.75,angle=0]{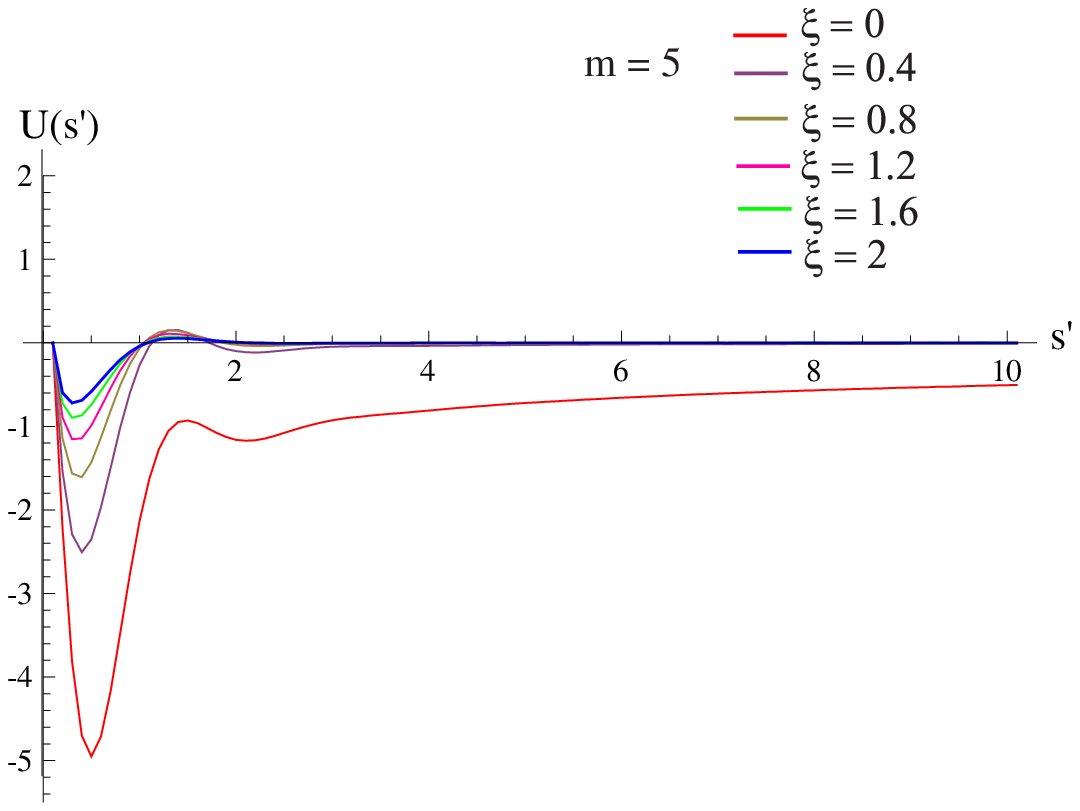}
   \caption{ Transverse (top) $m=1$ and (bottom) $m=5$ wakes with $\xi$ variable.}
   \label{fig:Fig9}
  \end{center}
 \end{figure}

The $m=1$ transverse wake is shown in Fig.~\ref{fig:Fig7} for the
first and second approximations (equivalent to Chao's formula and
Bane and Sands' respectively). With our formula we can study the
general case. Fig.~\ref{fig:Fig8} shows the different transverse
modes, for $\xi$=0, and Fig.~\ref{fig:Fig9} shows the dependence
on $\xi$ for $m=1$ and $m=5$.
\begin{figure}[!htb]
  \begin{center}
   \includegraphics[scale=0.75,angle=0]{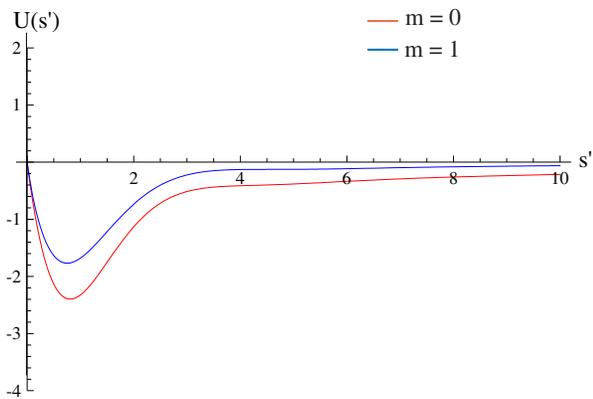}
   \caption{ The $m=0$ transverse wake, doubled,
   and the $m=1$ wake for $\xi=0.5$.}
   \label{fig:Fig10}
  \end{center}
 \end{figure}
The dominant longitudinal effect is the $m=0$ mode, and the
dominant transverse mode is for $m=1$. There is, as Chao makes
clear (\cite{Chao}, p81) no linkage between modes of different
orders to be obtained from the Panofsky-Wenzel theorem. However
the $m=0$ and $m=1$ longitudinal modes are connected by a simple
factor of ${2 \over b^2}$ as discussed in Section~\ref{sec:HOM}.
In the literature (e.g. \cite{Bane2} Equation 3) one finds the
expression
\begin{equation}
\tilde E_\perp^1 = {2 \over b^2 k} \tilde E_{||}^0
\end{equation}
This is exact for resistive wakefields only in the long wavelength
approximation. We investigate the validity of this widely-used
approximation in the general case. This difference is small, but
it is rather larger for the corresponding transverse wakes, as
shown in Fig.~\ref{fig:Fig10}.

\section{\label{sec:level4}AC Conductivity}

In the classical Drude model for the ac conductivity case, in
units of normalised wavenumber $K=s_0k$, we have
\begin{equation}\label{acsigma}
  \tilde{\sigma} = \frac{\sigma}{1-i K \Gamma}
\end{equation}
where we have introduced the dimensionless relaxation factor
$\Gamma = c \tau /s_0$. For a $1$ cm radius copper tube at room
temperature the relaxation time is $\tau=2.7\times 10^{-14}$s or
$c\tau=8.1$ $\mu$m giving  $\Gamma=0.4$, so we explore $\Gamma$
values in the range $0$ to $5$. In order to calculate the ac
resistive wall wakefield the dc conductivity $\sigma$ in
Equation~\ref{eq:third} is replaced by the general form of the ac
conductivity $ \tilde{\sigma}$. Hence,
\begin{equation}
  \lambda = \frac{b}{s_0^2}\sqrt{|K|}(1+K^2
  \Gamma^2)^{-1/4}[i\sqrt{1+t}
  \pm \sqrt{1-t}]
\end{equation}
while
\begin{equation}
  \frac{\lambda}{K} = \frac{b}{s_0}\frac{1}{\sqrt{|K|}(1+K^2
  \Gamma^2)^{1/4}}\Bigl[[\sqrt{1-t}\pm i\sqrt{1+t}\Bigr],
\end{equation}
with
\begin{equation}
  t = \frac{|K|\Gamma}{\sqrt{1+K^2 \Gamma^2}}.
\end{equation}

Following the same procedure employed on Section~\ref{sec:level2}
we express the inverse Fourier Transform of the m=0 mode of the
longitudinal component of the wakefield in terms of the even and
odd parts. The simplest of cases given by
Equation~\ref{eq:seventh} becomes,
\begin{eqnarray}
f_{even}(K)&=&-\frac{\sqrt{1-t}}{l}, \nonumber \\
f_{odd}(K)&=&\frac{i\sqrt{1+t}}{l}
\end{eqnarray}
where
\begin{equation}
  l=\frac{1}{\sqrt{|K|}(1+K^2 \Gamma^2)^{1/4}}.
\end{equation}

\begin{figure}[!htb]
  \begin{center}
   \includegraphics[scale=0.7,angle=0]{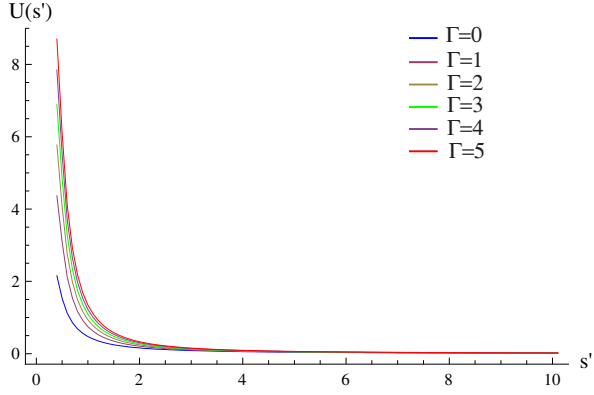}
   \caption{ The Long range wake for various values of $\Gamma$.}
   \label{fig:Fig11}
  \end{center}
 \end{figure}

The results are shown in Fig.~\ref{fig:Fig11} for various values
of $\Gamma$.
If we consider the more accurate formula given by
Equation~\ref{BandS} our technique gives
\begin{eqnarray}\label{BS_AC}
f_{even}(K)&=&-\frac{2
l\sqrt{1-t}}{\Bigl(\frac{K}{2}-l\sqrt{1+t}\Bigr)^2+l^2
(1-t)}, \nonumber \\
f_{odd}(K)&=&-
\frac{2i\Bigl(\frac{K}{2}-l\sqrt{1+t}\Bigr)}{\Bigl(\frac{K}{2}-l\sqrt{1+t}\Bigr)^2+l^2
(1-t)}
\end{eqnarray}

For the full version of the dc regime Equation~\ref{eq:first} the
odd and even parts have counterparts in the ac regime as
\begin{eqnarray}\label{full_AC}
f_{even}(K)=-8\,l^2 t (\xi^2\sqrt{1-t^2}+2\xi
lK\sqrt{1-t}\nonumber \\
+4l^3K\sqrt{1-t})/d(K) \nonumber \\
f_{odd}(K)= -8i\, l^2 t\Bigl[\xi^2
t-2\xi l(l-K\sqrt{1+t})\nonumber \\
-4l^3K\sqrt{1+t}+2l^2 K^2\Bigr]/d(K)
\end{eqnarray}
with the denominator given by
\begin{eqnarray}
d(K)=\Bigl(\xi^2\sqrt{1-t^2}+2\xi
lK\sqrt{1-t}+4l^3K\sqrt{1-t}\Bigr)^2 \nonumber \\
+\Bigl[\xi^2 t-2\xi l(l-K\sqrt{1+t})-4l^3K\sqrt{1+t}+2l^2
K^2\Bigr]^2 \nonumber
\end{eqnarray}

\begin{figure}[!htb]
  \begin{center}
    \includegraphics*[scale=0.55,angle=0]{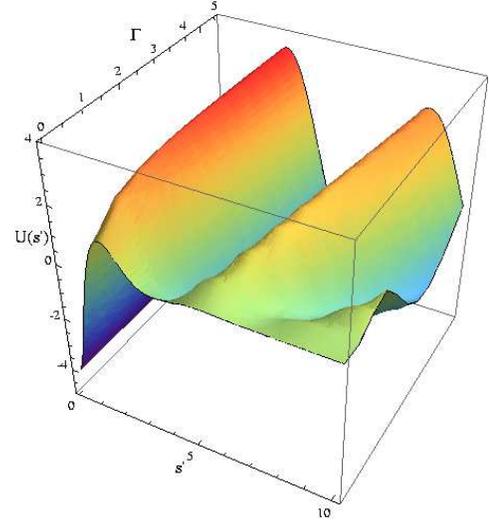}
    \includegraphics*[scale=0.75,angle=0]{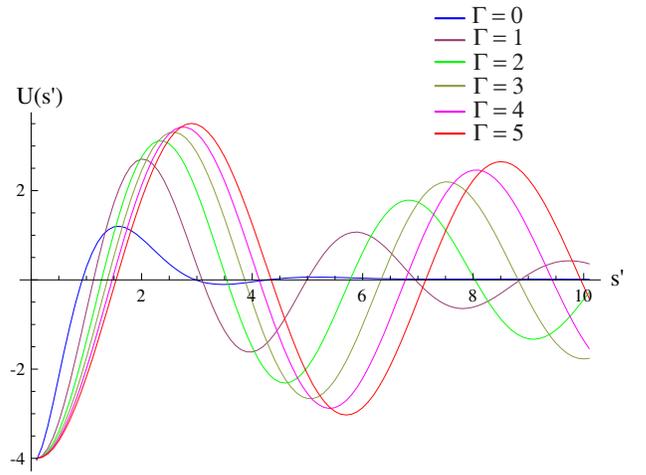}
   \caption{ The Longitudinal wake in the second approximation for various $\Gamma$ values.}
   \label{fig:Fig12}
  \end{center}
 \end{figure}

For $\xi=0$ one can see that Equation~\ref{full_AC} reduces to
Bane and Sands' approximation, Equation~\ref{BS_AC} as seen
before. Fig.~\ref{fig:Fig12} shows the results for various values
of $\Gamma$.
These wakes show a fairly strong dependence of $\Gamma$ and the
effects of AC conductivity may be important in a particular case.
However $\xi$ will always be very small so we include it only for
completeness and in practice one can probably use $\xi=0$ as a
good approximation.

For higher order modes the impedance is given by
Equation~\ref{fullm}. In the ac regime the odd and even parts are
\begin{eqnarray}
f_{even}(K)&=&-8\,l^2 t
(\xi^2\sqrt{1-t^2}+2\xi lK\sqrt{1-t}\nonumber \\
& &+4l^3K\sqrt{1-t})/d(K)\nonumber \\
f_{odd}(K)&=&-8i\,l^2 t\Bigl[\xi^2 t-2\xi
l(l-K\sqrt{1+t})\\
& &-4l^3K\sqrt{1+t}+4l^2
\Bigl(\frac{K^2}{m+1}-m\xi\Bigr)\Bigr]/d(K)\nonumber
\end{eqnarray}
with the denominator given by
\begin{eqnarray}
d(K)&=&\Bigl(\xi^2\sqrt{1-t^2}+2\xi
lK\sqrt{1-t}+4l^3K\sqrt{1-t}\Bigr)^2 \nonumber \\
& &+\Bigl[\xi^2 t-2\xi
l(l-K\sqrt{1+t})-4l^3K\sqrt{1+t}\nonumber \\
& &+4l^2\Bigl(\frac{K^2}{m+1}-m\xi\Bigr) K^2\Bigr]^2 \nonumber
\end{eqnarray}
\begin{figure}[!htb]
  \begin{center}
    \includegraphics*[scale=0.75,angle=0]{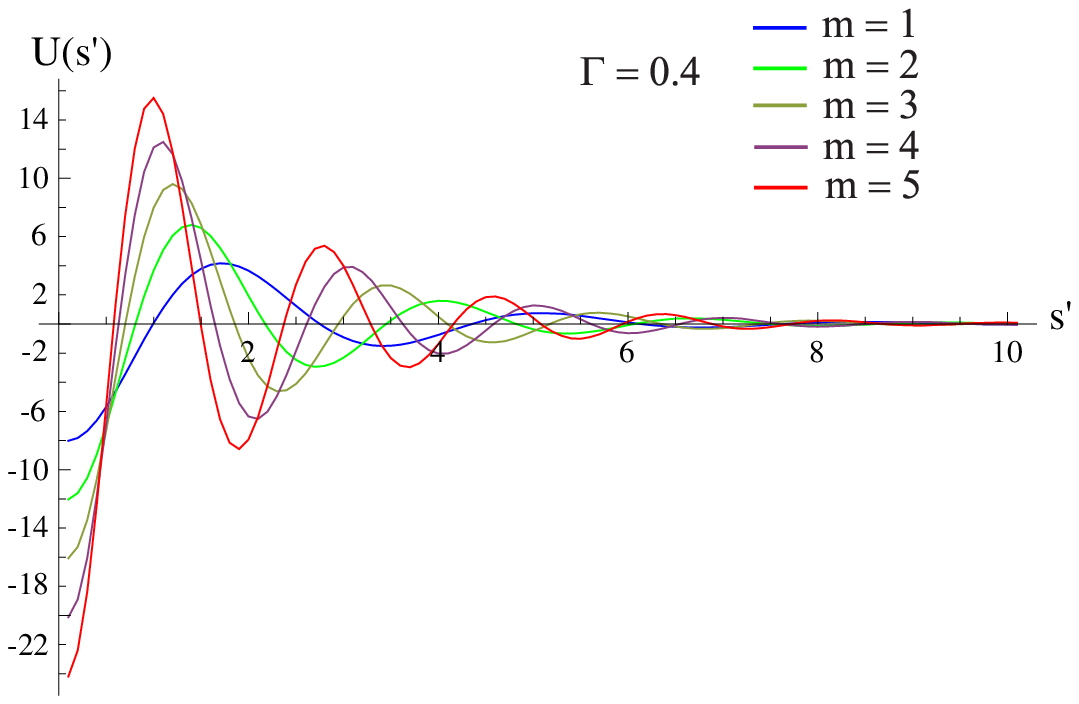}
    \includegraphics*[scale=0.75,angle=0]{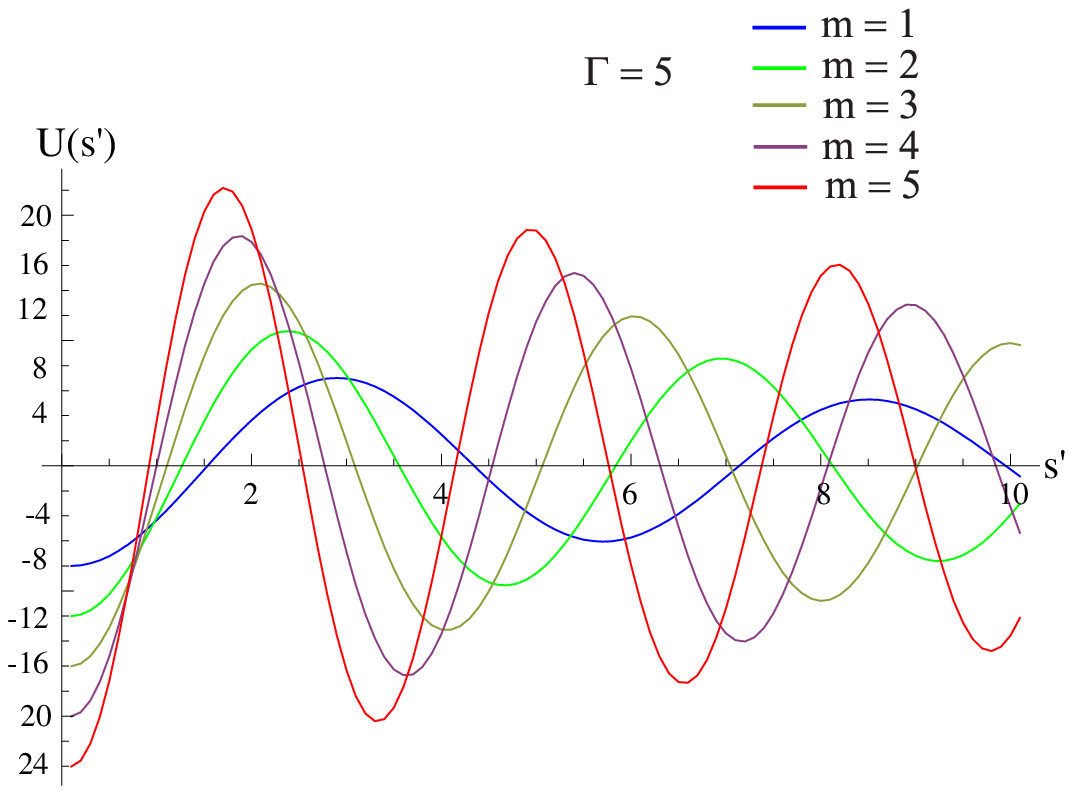}
   \caption{ The longitudinal wake for
various modes ($m>0$) for $\Gamma=0.4$ (top) and $\Gamma=5$
(bottom).}
   \label{fig:Fig13}
  \end{center}
 \end{figure}
 \begin{figure}[!htb]
  \begin{center}
    \includegraphics*[scale=0.55,angle=0]{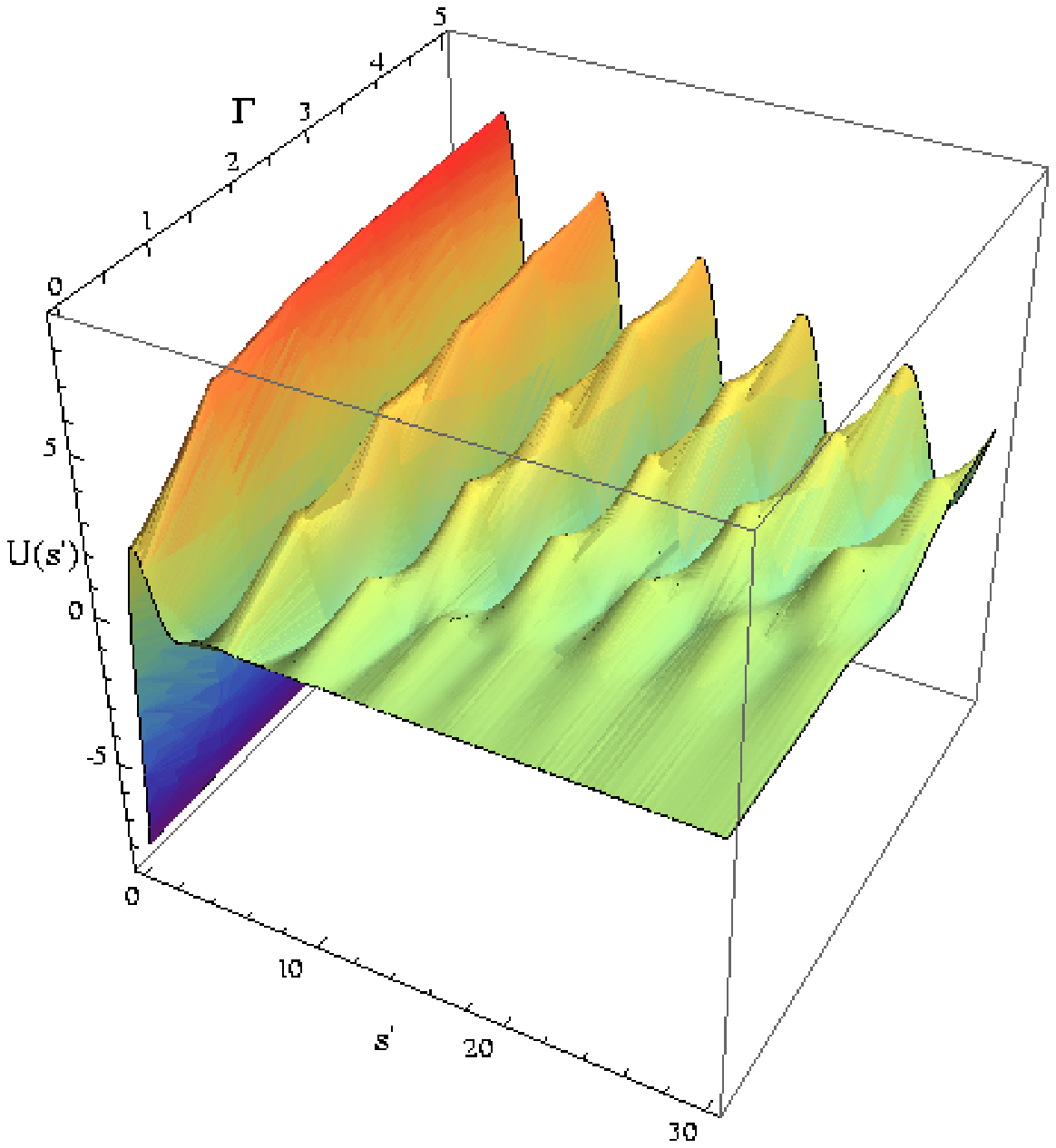}
    \includegraphics*[scale=0.55,angle=0]{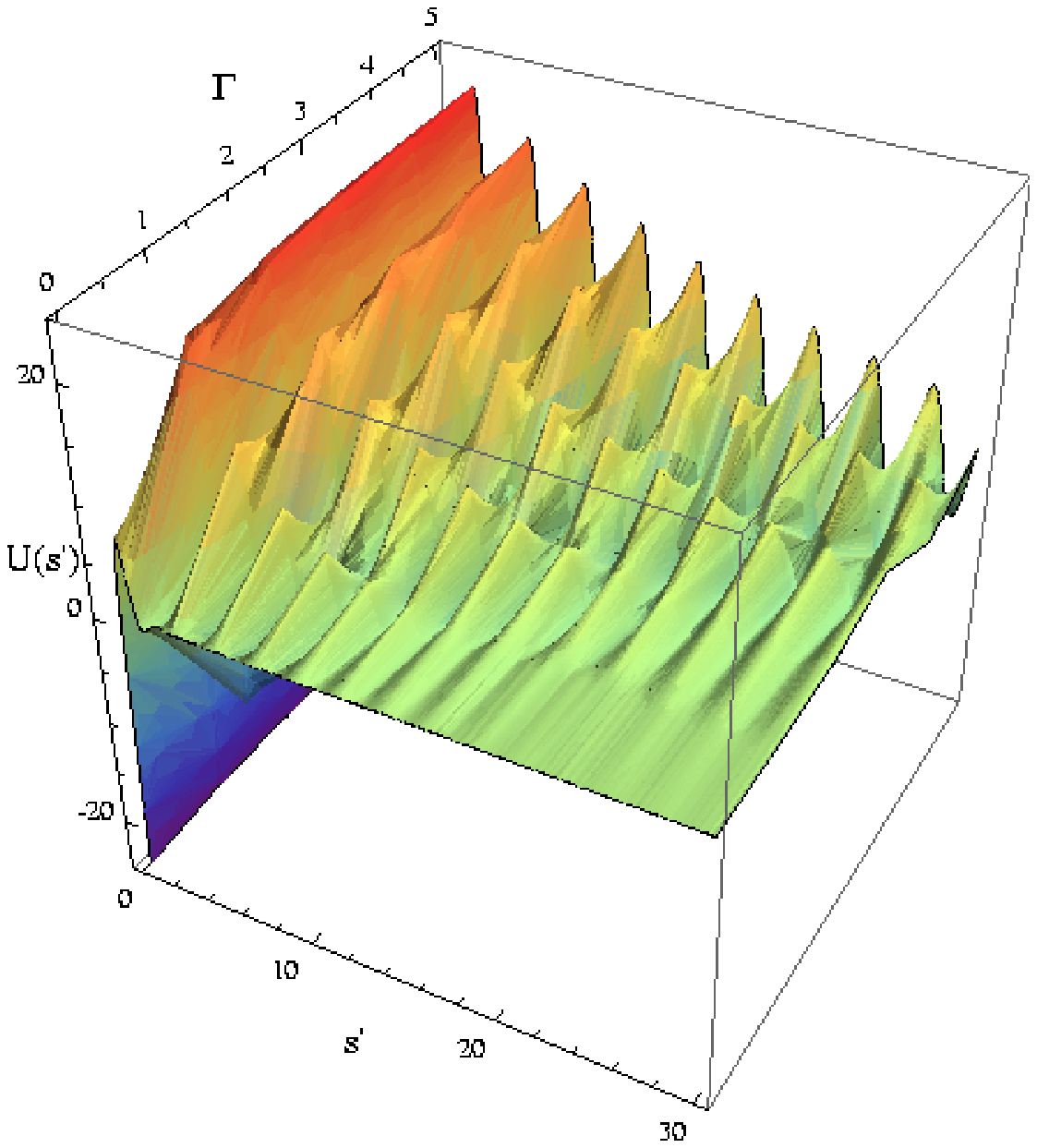}
   \caption{ The m=1 (top) and m=5 (bottom) longitudinal wake as a function of $\Gamma$.}
   \label{fig:Fig14}
  \end{center}
 \end{figure}
In Fig.~\ref{fig:Fig13} we show the different modes dependance for
two particular values of $\Gamma$, $0.4$ and $5.0$ and $\xi=0$.

\begin{figure}[!htb]
  \begin{center}
    \includegraphics*[scale=0.75,angle=0]{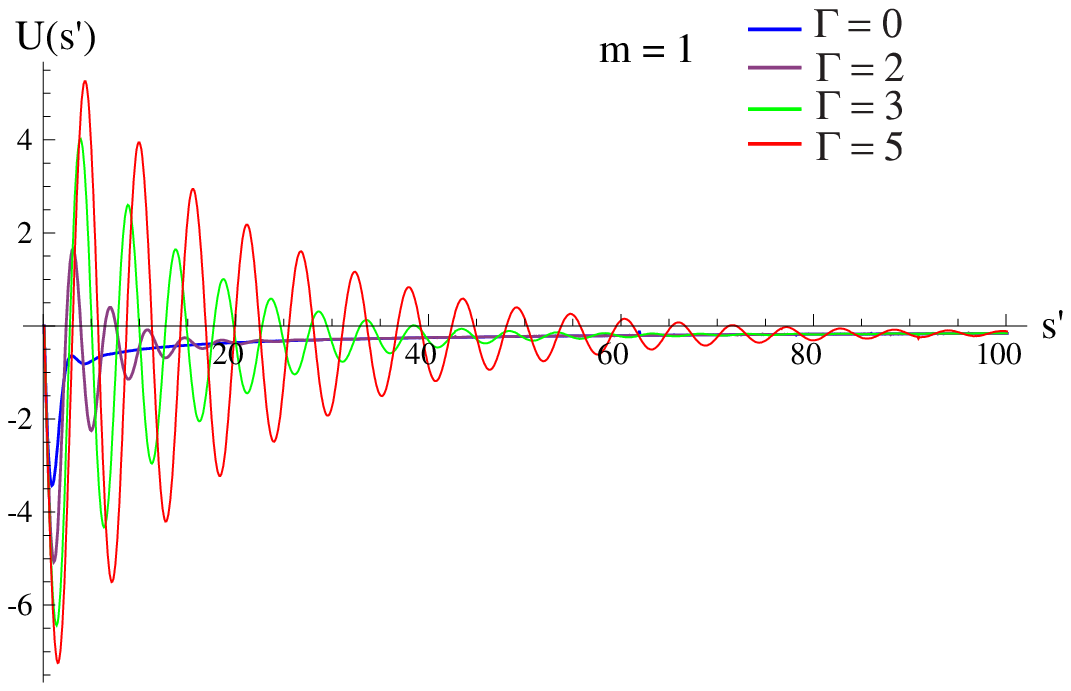}
    \includegraphics*[scale=0.75,angle=0]{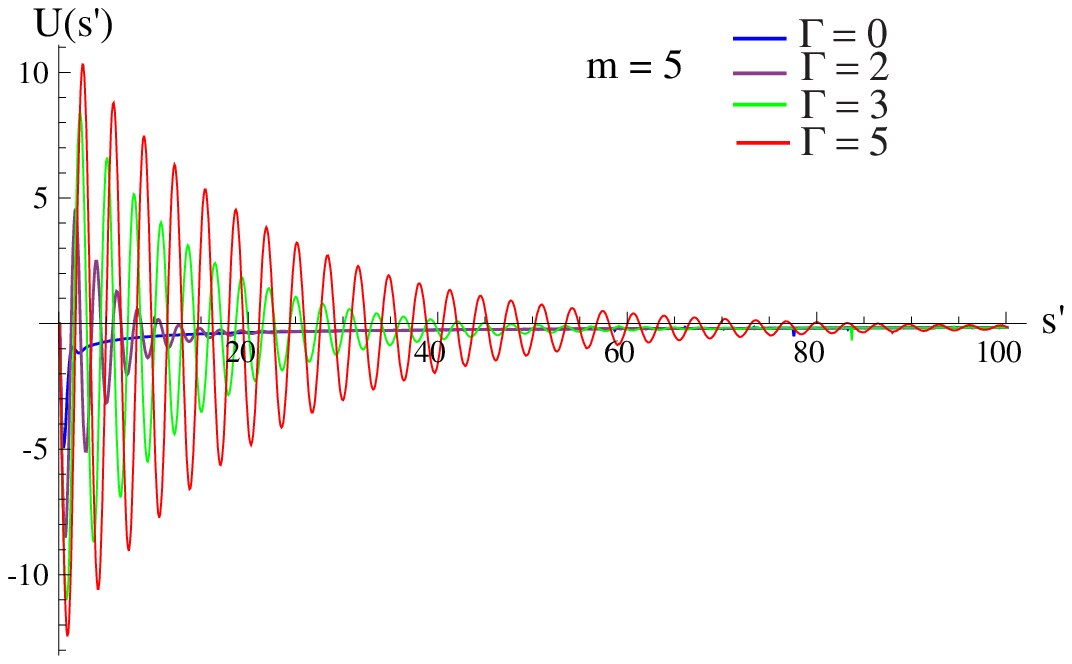}
   \caption{ The m=1 (top) and m=5 (bottom) transversal wake for various values of $\Gamma$, all for $\xi$=0.}
   \label{fig:Fig15}
  \end{center}
 \end{figure}

The $\Gamma$ dependence of two modes, mode $m=1$ and $m=5$ is
shown in Fig.~\ref{fig:Fig14} also for $\xi=0$. The equivalent
transverse wake is shown in Fig.~\ref{fig:Fig15}. It is clear that
for all modes, increasing $\Gamma$ increases the size of the wake
effect, both longitudinal and transverse over the same distance.
For the higher modes the effect is stronger.

\section {Implementation}
The integrals used to generate the plots in this paper were
performed using Mathematica~\cite{Mathematica}. Tables with $6
\times 1001 \times 21$ elements were written to file, covering the
ranges $0 \leq \Gamma \leq 5$, $0 \leq s' \leq 100$ and $0 \leq
\xi \leq 2$ respectively. A separate table was used for each mode,
and tables were written separately for longitudinal and transverse
wakes.

For ease of use we have defined a small object {\tt
collimatortable} written in {\tt C++} but portable between Merlin,
Placet, and hopefully other simulation codes. When created by a
call to the constructor {\tt collimatortable(filename, Gamma, xi)}
it will read the full 3 dimensional table from {\tt filename} and
construct a one dimensional table, using parabolic interpolation
between the 9 closest points, for $s'$ at this value of $\Gamma$
and of $\xi$, both of which default to zero. It does this because
when tracking particles through a collimator the values of $s$ (or
$s'$) are different for each pair of particles (or slices),
whereas the values of $\Gamma$ and $\xi$ are the same, as they
depend on the properties of the collimator, not the bunch. Then
the one dimensional table can be used by the member function {\tt
collimatortable::interpolate(sprime)} to find the value for any
value of $s'$ in the range.

These data and program files are obtainable from the authors
\cite{ourwebsite}.

\subsection{MERLIN}
The MERLIN program~\cite{Merlin} contains a general wake formalism
which has been extended~\cite{ourmerlin} for high order modes
(though only for axially symmetric apertures) and it is easy to
incorporate these wakefields. A class {\tt ResistivePotentials}
which inherits from {\tt SpoilerWakePotentials} calculates
$\Gamma$ and $\xi$ from the dimensions and properties given and
uses {\tt collimatortable} to read the files;  functions {\tt
Wtrans(z,m)} and {\tt Wlong(z,m)} use the {\tt interpolate}
function to get the wake experienced by the particles.

For efficiency, the bunch is divided longitudinally into a number
of slices $N_s$ (typically $100$) and each wake function is called
only $N_s(N_s-1)/2$ times.

\subsection{PLACET}
The PLACET program~\cite{Placet} contains only the lowest order
$m=1$ transverse mode, though it does include the effect of
rectangular apertures using as an Ansatz the Yokoya factors
$y_{eff} = 0.822 y_1+ 0.411 y_2$, where $y_1$ and $y_2$ are the
displacements of the leading and trailing particle. $y_{eff}$ is
then used in the formula for circular apertures.

In the current implementation a complicated system switches
between different formulae depending on the sizes of the pipe and
bunch. The integral resulting from the analytic transform of
Equation~\ref{BandS} is calculated numerically. We replace these
with a single call to the {\tt interpolate} function. $100$ lines
of code in the original are replaced by $21$ lines in the new
version. The wake function is called for each slice for each
macroparticle, $~N_s N_p/2 $ times.


\section{\label{sec:example}Examples}
\subsection {Simple collimator}
As a numerical example we consider a $1$ m long collimator made of
Titanium, with conductivity $\sigma=2.33\times10^6\,(\Omega
m)^{-1}$. The radius is $1.4$ mm and the relaxation time is
$2.7\times 10^{-14}\ $s. This gives $s_0=0.165\,\mu$m, $\Gamma=
0.49$ and $\xi=0.00014$.
\begin{figure}[!htb]
  \begin{center}
    \includegraphics*[scale=0.35,angle=0]{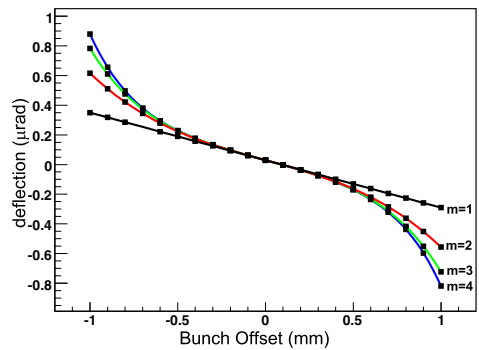}
   \caption{ Merlin prediction for the deflection as a function of offset.}
   \label{fig:Fig16}
  \end{center}
 \end{figure}
\begin{figure}[!htb]
  \begin{center}
    \includegraphics*[scale=0.35,angle=0]{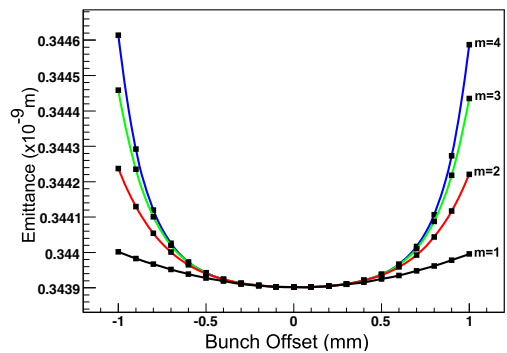}
   \caption{ Merlin prediction for emittance as a function of offset.}
   \label{fig:Fig17}
  \end{center}
 \end{figure}

This collimator and the beam bunch properties are those of the
recent tests at SLAC End Station A~\cite{ESA}, with
$\beta_x=128.2$ m, $\beta_y=11.9$ m, $\epsilon_x=5.49\times
10^{-9}$ m, $\epsilon_y=3.44\times10^{-10}$ m, initial beam energy
$p=28.5$ GeV and bunch length $\sigma_z=0.3$ mm, with the sole
difference of $\epsilon_x=\epsilon_y=3.44\times10^{-10}$ m when
using Merlin. The bunch of $N_p=10^{10}$ particles was modelled by
$50000$ macroparticles in both simulations.

\begin{figure}[!htb]
  \begin{center}
    \includegraphics*[scale=0.65,angle=0]{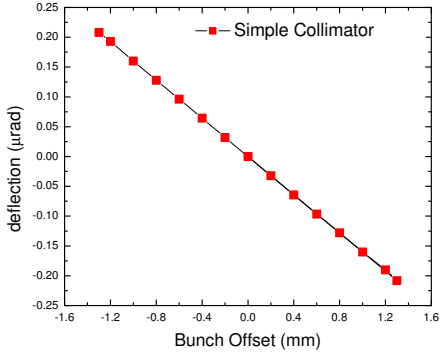}
   \caption{PLACET prediction for the deflection as a function of offset.}
   \label{fig:Fig18}
  \end{center}
 \end{figure}

\begin{figure}[!htb]
  \begin{center}
    \includegraphics*[scale=0.65,angle=0]{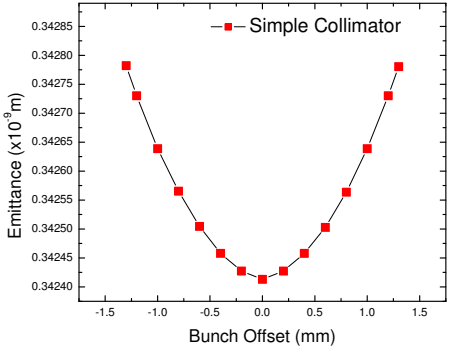}
   \caption{ PLACET prediction for emittance as a function of offset.}
   \label{fig:Fig19}
  \end{center}
 \end{figure}

We show the predictions of Merlin (PLACET) for the mean kick of
the bunch in Fig.~\ref{fig:Fig16} (Fig.~\ref{fig:Fig18}) and for
the increased geometric emittance in Fig.~\ref{fig:Fig17}
(Fig.~\ref{fig:Fig19}), respectively. The total emittance increase
leading to the loss of luminosity at the IP caused by wakefields
is due in part to jitter amplification and in part to the
increased size of the bunch itself~\cite{Zimmermann}. This latter
effect is what we show. If one considers a bunch as a series of
slices it is clear that different slices receive different angular
kicks, the first slice receiving no kick at all, and this
difference between slices increases the angular size of the bunch
as a whole.

The deflection of Figs.~\ref{fig:Fig16} and~\ref{fig:Fig18}
compounded with transverse jitter, may also contribute to the loss
of luminosity by an amount that depends on the betatron phases at
the collimator and at the IP. In previous analyses, such
as~\cite{PT},~\cite{EPACpaper} it was assumed that the incoherent
intrinsic increase of emittance is $0.4$ times coherent term
produced by jitter. Here we show that it can be calculated
explicitly.

The results from Merlin and PLACET agree surprisingly well, given
that one uses the Yokoya terms and the other does not. The overall
shift in the emittance of Figs.~\ref{fig:Fig17}
and~\ref{fig:Fig19} is due to the different random bunches and
using only $50000$ macro particles. Merlin results show that
higher modes become important (only) for offset $\geq b/2$.

\subsection {Extreme collimator}
We consider a $1$ m long collimator made of graphite, with
conductivity $\sigma=7.14\times 10^4(\Omega m)^{-1}$ and very
narrow pipe of radius $0.125$ mm. This gives $s_0=0.105\,\mu$m,
$\Gamma=2.85$ and $\xi=0.007$. For comparison purposes we consider
a beam bunch of the same properties as for the previous
collimator, but with the emittance ten times smaller, while the
displacements of the beam from the central axis are between $0$
and $0.115$ mm.
\begin{figure}[!htb]
  \begin{center}
    \includegraphics*[scale=0.35,angle=0]{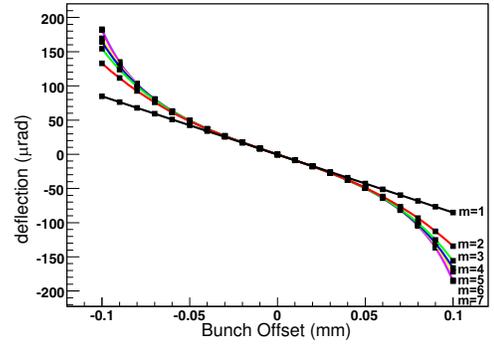}
   \caption{Merlin prediction for the deflection as a function of offset.}
   \label{fig:Fig20}
  \end{center}
 \end{figure}
\begin{figure}[!htb]
  \begin{center}
    \includegraphics*[scale=0.35,angle=0]{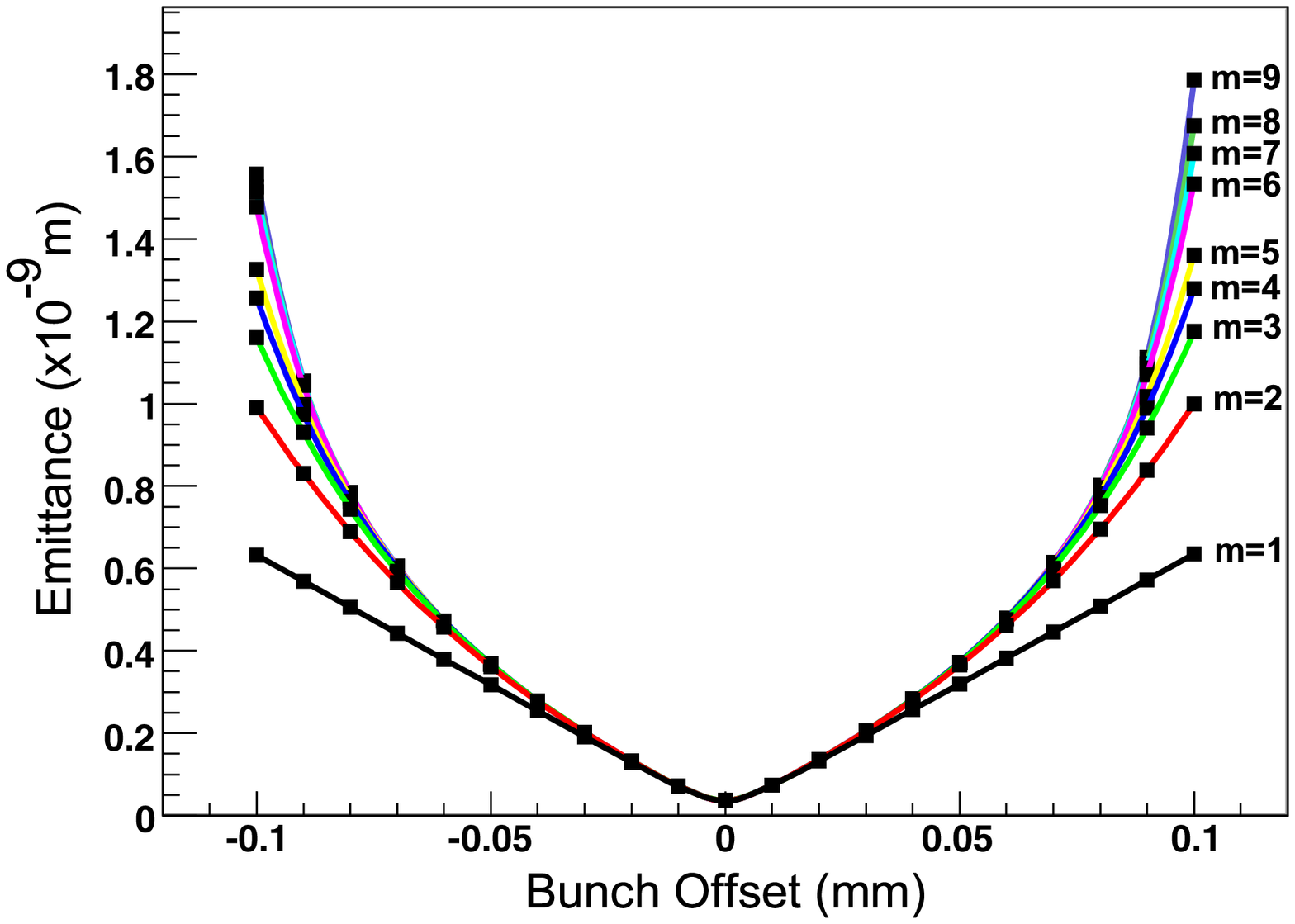}
   \caption{Merlin prediction for emittance as a function of offset.}
   \label{fig:Fig21}
  \end{center}
 \end{figure}
\begin{figure}[!htb]
  \begin{center}
    \includegraphics*[scale=0.65,angle=0]{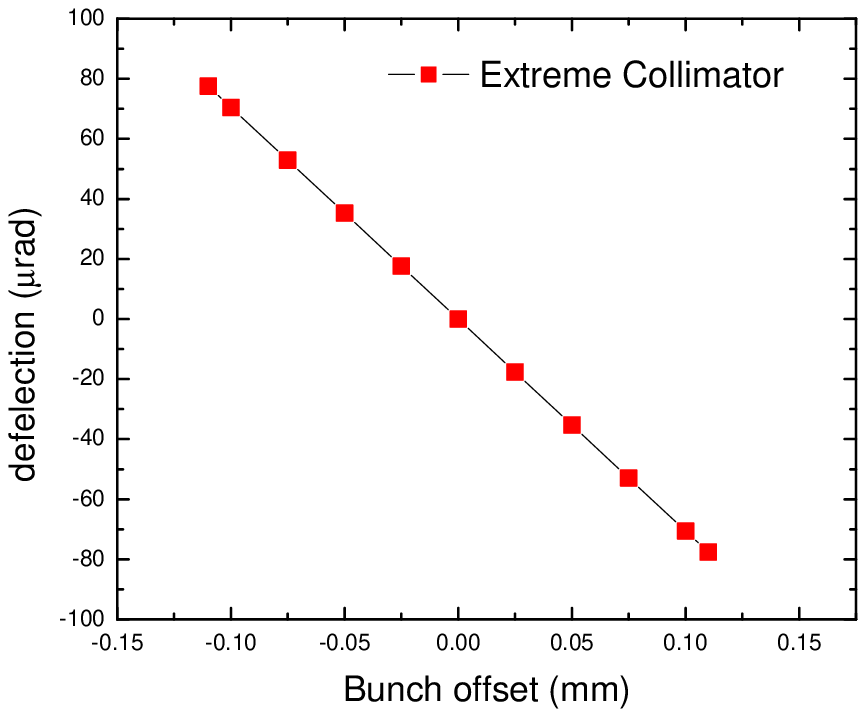}
   \caption{PLACET prediction for the deflection as a function of offset.}
   \label{fig:Fig22}
  \end{center}
 \end{figure}
\begin{figure}[!htb]
  \begin{center}
    \includegraphics*[scale=0.65,angle=0]{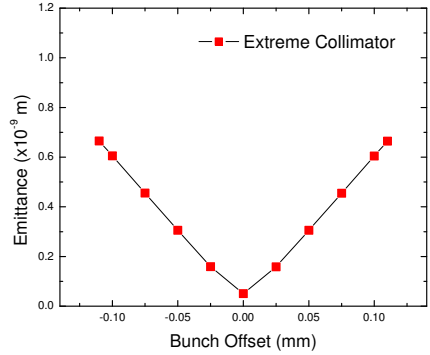}
   \caption{ PLACET prediction for emittance as a function of offset.}
   \label{fig:Fig23}
  \end{center}
 \end{figure}

Again, the agreement is good (Figs.~\ref{fig:Fig20}
-~\ref{fig:Fig23}). Merlin shows that higher modes become
significant sooner than they do for the previous case. Also it can
be seen that more modes are needed for convergence when the bunch
displacement is large compared to pipe radius.

\section{Conclusions}
Longitudinal and transverse resistive wakes can be calculated in
simulations programs with a technique applicable for any general
particle separation using pretabulated numerical Fourier
Transforms and implemented in a way which is computationally
efficient. Higher order angular modes can easily be included, and
so can be AC conductivity effects. We have shown that this can be
used by the Merlin, PLACET programs and other codes can be added
in due curse. Results are given for two types of collimator, and
show that the effects are as expected. This can be used in
simulations of full beamline lattice for accelerators such as CLIC
and the LHC.
\newpage 
\bibliography{apssamp8}

\end{document}